\documentclass{aa}
\usepackage{amsmath}
\usepackage{amssymb}
\usepackage[latin1]{inputenc}
\usepackage{graphicx}
\usepackage{natbib}
\usepackage{txfonts}

\begin{document}

\title{Dust-scattered X--ray halos around two Swift gamma-ray bursts: GRB 061019 and GRB 070129}

  \author{G. Vianello\inst{1,2} \and A. Tiengo\inst{1} \and S. Mereghetti\inst{1}}

   \offprints{G. Vianello, email: vianello@iasf-milano.inaf.it}

  \institute{INAF - Istituto di Astrofisica Spaziale e Fisica Cosmica Milano,
          Via Bassini 15, I-20133 Milano, Italy
          \and
              Dipartimento di Fisica e Matematica, Universit\`{a} dell'Insubria,
              via Valleggio 11, I-22100 Como, Italy
   }

  \date{}

\abstract{Two new expanding X--ray rings were detected by the
Swift XRT instrument during early follow-up observations of \object{GRB
061019} and \object{GRB 070129}, increasing to 5 the number of dust
scattering X--ray halos observed around GRBs. Although these two
halos were particularly faint, a sensitive analysis can be
performed that optimizes the method originally developed
by Tiengo \& Mereghetti (2006) to analyze dust scattering rings
observed with XMM-Newton for the Swift satellite. In the case of \object{GRB 061019}, a known giant
molecular cloud is identified as the one responsible for the
scattering process, and its distance is accurately measured
(d=940$\pm$40 pc) through the  dynamics of the expanding ring. In
the second case, XRT observed both the main peak of the prompt
emission of \object{GRB 070129} and the scattering halo, but the small
number of detected halo photons prevents us from distinguish between different dust
models. \\
 }

\authorrunning{G.Vianello, A. Tiengo \& S. Mereghetti}
\titlerunning{GRB halos}

\maketitle

%

\section{Introduction}

The scattering of X--rays from interstellar dust grains is often
responsible for the formation of diffuse halos surrounding bright
sources. The study of such halos can then be used to infer the properties
of the dust and its spatial distribution \citep{ML91,D03}, and, in case
of variable sources, can also yield information on the source's distance \citep{ts73}.

Thanks to rapid follow-up observations with sensitive X--ray
telescopes, it has recently become possible to observe halos
around gamma-ray bursts. Scattering halos due to dust in our
Galaxy have been detected around three gamma-ray bursts: GRB
031203 \citep{vaughan04},  GRB 050724 \citep{vaughan06}, and GRB
050713A \citep{tm}. Due to the short duration of the bursts and the
relatively small thickness of the dust layers, such halos appear
as expanding rings. For the dust distances $D_{dust}$ typically involved here (hundreds of parsecs) the rings have angular radii
$\theta$ of a few arcminutes after several thousand seconds from
burst time:
\begin{equation}
\theta[arcsec]=\sqrt{\frac{827}{D_{dust}[pc]}t[s]}\mbox{\qquad .}
\label{eq:key_formula}
\end{equation}

The brightness of the halo, I$_{HALO}$,  depends on the intensity
of the GRB emission, I$_{GRB}$, and on the scattering optical
depth, $\tau$: I$_{HALO}$ = I$_{GRB}$ (1--e$^{-\tau}$) $\simeq$
$\tau$ I$_{GRB}$. The same dust is responsible both for the X--ray halos and for the interstellar reddening, so $\tau$ is usually inferred from measurements of the optical reddening $A_V$. However, different relations have been reported
in the literature between $A_V$ and $\tau$. Draine has  developed a detailed physical model of the dust \citep{D03} and extracted from this model the following relation between optical extinction and scattering optical depth in the soft X--ray range \citep{DB04}:
\begin{equation}
\tau(E)\approx 0.16 A_V \left(\frac{E}{1 \mbox{keV}}\right)^{-1.8}\mbox{\qquad .}
\label{eq:draine}
\end{equation}
However, a different relation was empirically found by \citet{PS95}, that is:
\begin{equation}
\tau(E)\approx 0.056 A_V \left(\frac{E}{1 \mbox{keV}}\right)^{-2}\mbox{\qquad .}
\label{eq:predehl}
\end{equation}
If both I$_{HALO}$ and I$_{GRB}$ are known or
can be estimated, as in the case of GRB 050724 \citep{vaughan06},
some information on the amount of dust and its properties can be
derived. Otherwise, by knowing or making assumptions about the amount
of dust and its properties, the measurement of the radiation
scattered in the halo can provide some information on the GRB
prompt emission at a few keV. This has been done, for example, for GRB
031203 \citep{vaughan04,tm}. 
Here we report on two new dust-scattering rings recently observed
with Swift around the gamma-ray bursts \object{GRB 061019} \citep{GCN5737}
and \object{GRB 070129} \citep{GCN6065}. Our analysis is based on a method
particularly convenient in the case of faint, expanding halos
\citep{tm}, which is briefly described in the next section.

\section{Analysis method}

The method developed by \citet{tm}
for detecting and analyzing expanding dust halos can be
applied to any imaging X--ray detector. We applied it to data
obtained with the X--Ray Telescope (XRT) onboard Swift
\citep{xrt}.

For each count detected by the XRT, we computed the arrival time
$t$, in seconds from the burst time,  and its squared angular
distance from the GRB position, $\theta^2$. Based on these
quantities, we built a \textit{dynamical image}, that is a 2-D
histogram with $t$ on the x-axis and $\theta^2$ on the y-axis. In
such an image an expanding ring forms a straight line with an angular
coefficient depending on the dust distance, as can be seen from
Eq.\ref{eq:key_formula}. In this way we can easily detect a dust-scattering ring and estimate the distance of the dust layer from
the line slope. In order to determine the halo intensity and dust
layer distance and thickness, we removed the
GRB afterglow and any other point sources from the event list, and we computed the value $D=827 t/\theta^2$ for each detected event. In the ideal case
(thin dust layer and perfect instrument), every halo event will
have the same value of $D$, corresponding to the actual distance
of the dust layer. In real cases, instead, the instrument point
spread function (PSF) and the intrinsic width of the scattering dust
layer produce a distribution of $D$ centered on the dust layer
distance. If we plot the distribution of the quantity $D$
(\textit{dust distance distribution} hereinafter), we obtain a
single peak centered on the real dust layer distance, formed by
the photons belonging to the ring, superimposed on a background
distribution.

If the instrument PSF is described well by a
King function (as in the case of both XMM-EPIC and Swift-XRT), the
dust ring peak is well-modeled by a Lorentzian function. When the
observation is continuous (as typically in the case of XMM-EPIC
observations), the background distribution is a  power law, so the
dust distance distribution is described by a Lorentzian plus
a power law. Unfortunately, the Swift observations are always discontinuous, owing to the low earth orbit of the satellite and to
the particular observation plan adopted by XRT \footnote{Typically
the observing plan of a GRB afterglow is composed of several
individual exposures separated by time gaps.}. Since in this case the background distribution cannot be easily
described analytically, we developed a Montecarlo simulation
that reproduces the contribution to the  dust distance
distribution produced by the background counts, as well as the
dust scattering ring, taking the actual observation
plan into account. This simulation is also useful for understaning some
other aspects of the phenomenon better, such as the effects
due to the instrumental PSF and effective area or due to the dust
layer intrinsic width. Our simulation can generate a GRB prompt
X--ray emission based on input parameters describing its timing
and spectral characteristics; alternatively, it can use as
input an event file obtained during real observations of the
X--ray prompt emission. Then it models the scattering process
using the layer distance and thickness as input. 
Our simulation can use either of the relations (\ref{eq:draine}) or (\ref{eq:predehl}). Taking the characteristics of the instrument
(PSF and effective area), the background and the given observation
plan into account, the program generates an output event file in a standard
format allowing its analysis as that of a real observation.

\section{GRB 061019}

\object{GRB 061019} was discovered with the Swift Burst Alert Telescope
(BAT) and localized at Galactic coordinates l=$181.74^{\circ}$ and
b=$4.26^{\circ}$  \citep{GCN5728,GCN5733}. The XRT and UVOT began
their observations of the burst location 2800 s after the trigger
time $T_0$ = 04:19:06 UT. A dust scattering ring was
discovered \citep{GCN5737} in the first three orbits of XRT Photon
Counting data (from $\sim T_0+2800$ s to $\sim T_0+17000$ s).

\begin{figure}[tb]
    \centering
    \includegraphics[width=0.4\textwidth]{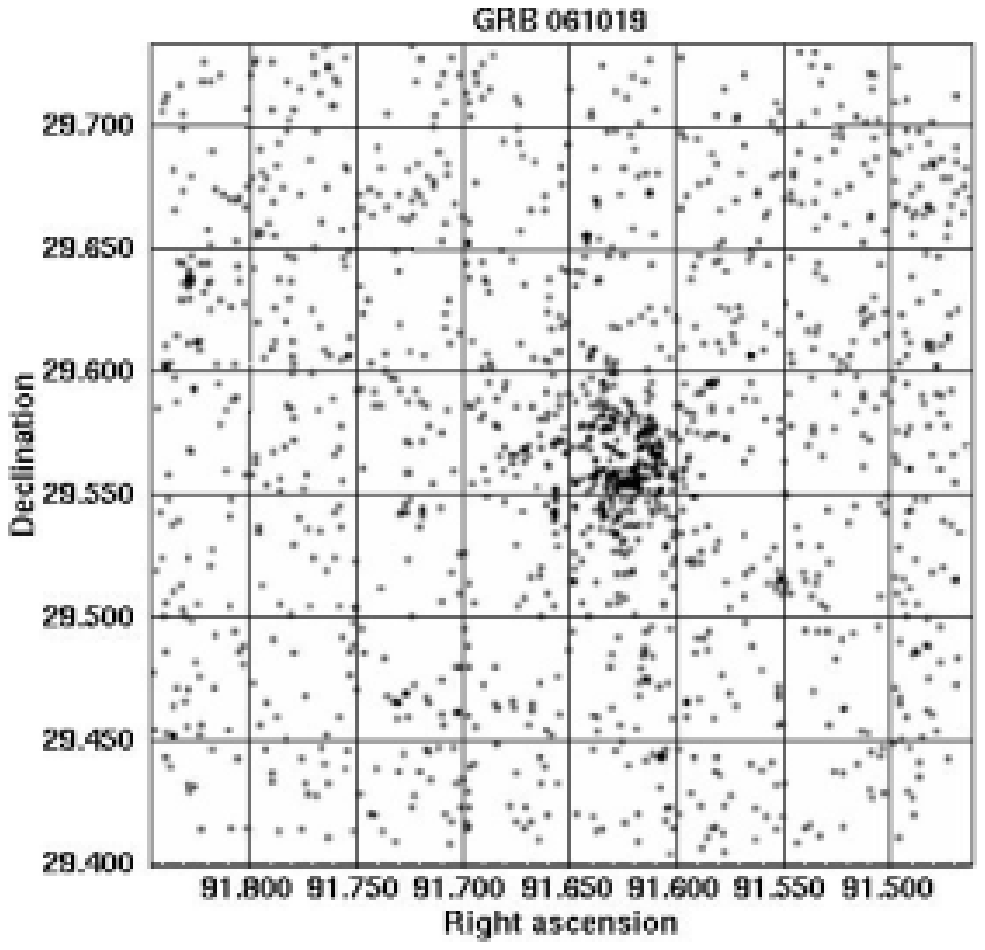}
    \includegraphics[width=0.5\textwidth]{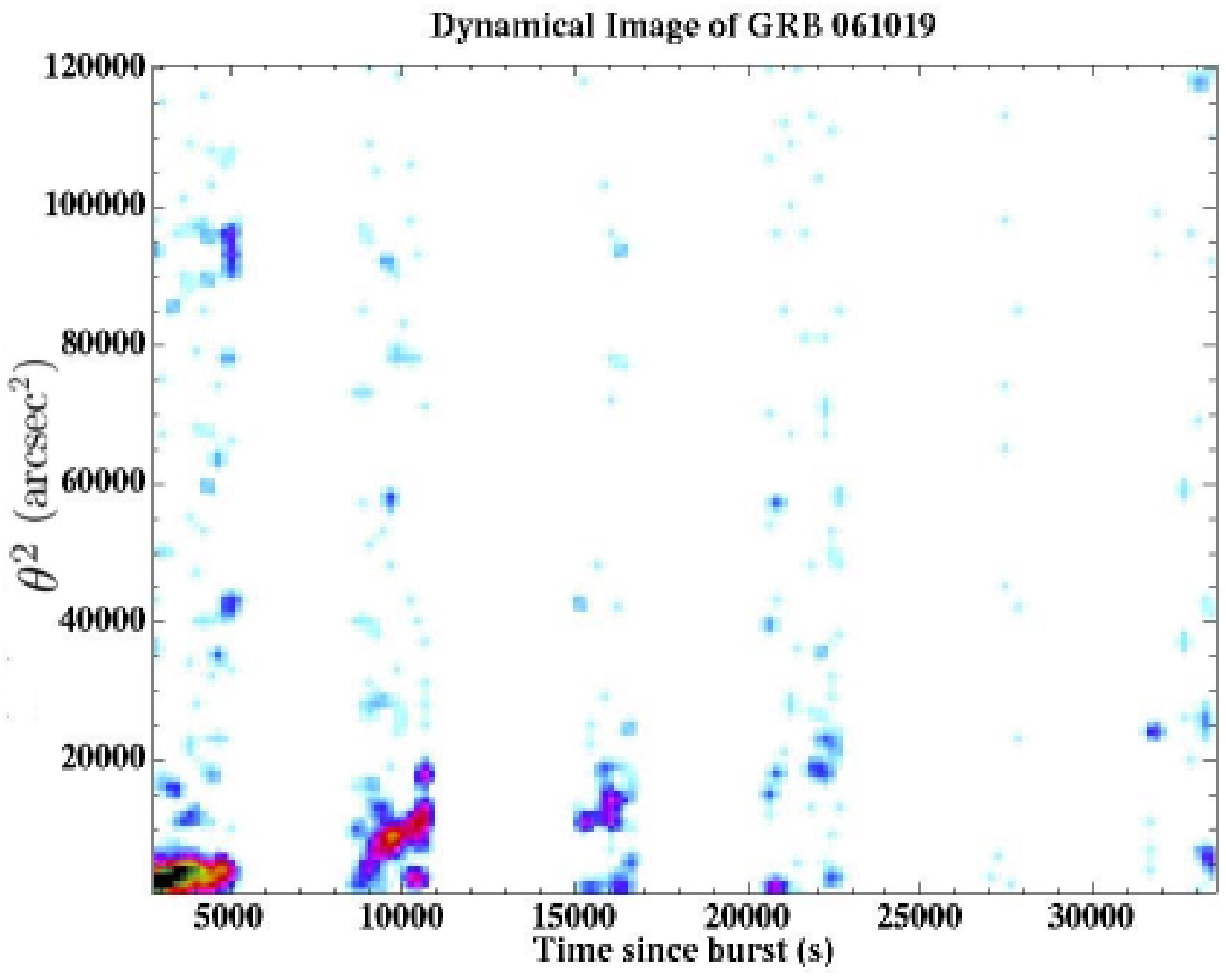}
    \caption{Image (top panel) and dynamical image (bottom panel) of \object{GRB 061019}
in the 0.2-4 keV energy range. We have excluded from the images a
circle with a radius of 30$''$ centered at the GRB position to
remove the afterglow contribution. In the upper panel the cross marks the afterglow position.}
    \label{fig:dyn_061019}
\end{figure}

\begin{figure}[tb]
    \centering
    \includegraphics[width=0.4\textwidth]{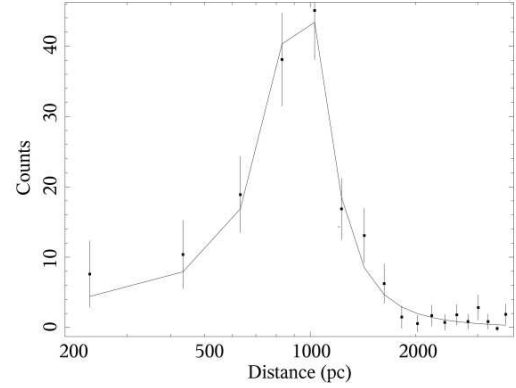}
    \caption{Background-subtracted dust distance
    distribution of \object{GRB 061019}. The peak is due to the expanding ring.}
    \label{fig:ddp061019}
\end{figure}

Our analysis of these data has confirmed the detection of the
ring. The scattering halo is very visible in the sky  image and in
the dynamical image shown in Fig.\ref{fig:dyn_061019}. These
images, as well as all the analysis discussed below, refer to the
0.2--4 keV energy range.
The line formed by the halo events in the dynamical image is rather wide due to  the
intrinsic thickness of the expanding ring, which  remains clearly
visible until $\sim 23000$ s after the burst.
We  estimated the background count rate from four regions of the
XRT field of view, at least $4^{\prime}$ away from the afterglow
position and with no evident point sources. The average count rate
was $(3.4\pm0.3)\times10^{-8}$ counts s$^{-1}$ arcsec$^{-2}$ and
no significant spatial variability was found by comparing the
different regions. Using the Montecarlo simulator described above
with this background rate (and its error), we estimated the dust 
distance distribution expected for the background counts. This
distribution was subtracted from the data, taking both the statistical (poissonian) errors and that of the
background rate into account, in order to derive the net distance distribution
for the halo shown in Fig.~\ref{fig:ddp061019}. After removing the
background, the halo counts are $\sim 180\pm20$ in the $0.2-4$ keV
energy range. The peak is well-fitted with a Lorentzian centered
at $D=941^{+43}_{-47}$ pc and with  FWHM width $\Delta
D=427^{+115}_{-99}$ pc ($\chi^2\mbox{/d.o.f.}=0.74$ with 16 d.o.f.). Our
simulations show that in this case the XRT PSF
produces a peak width of $\sim180$ pc. By fixing the Lorentzian
FWHM at this value, we obtained a fit to the peak with the same
center ($D=938\pm12$ pc), but with a significantly worse
chi-square ($\chi^2\mbox{/d.o.f.}=1.71$ with 17 d.o.f.) corresponing to a chance probability of 4\%, so that the hypothesis of a zero-thickness cloud is rejected at 3 sigma level. We thus conclude that
the PSF is only partially responsible for the observed ring width
and that the dust layer causing the scattering has an intrinsic
width $\Delta D_{intr}\gtrsim150$ pc.

The halo spectrum can be fitted with an absorbed power law. Due to
the small number of counts we fixed the hydrogen column
density $N_H$ to the best-fit value obtained for the afterglow
($N_H=0.9\times10^{22}$ cm$^{-2}$, \citealt{GCN5733}), finding a
photon index $\Gamma = 1.8\pm0.9$  and a halo fluence of
$\sim3\times10^{-9}$ cm$^{-2}$ ($\chi^2\mbox{/d.o.f.}=1.3$ with 2 d.o.f). Adopting the
optical depth-energy relation in Eq.\ref{eq:draine} and assuming the optical extinction reported in \citet{Schlegel} for the bust direction ($A_V=3.4$), we estimated the prompt X--ray fluence following the method described in \citet{tm} as 
$\sim2\times10^{-7}$ erg cm$^{-2}$ ($1-2$ keV) . If we
extrapolate the BAT spectrum ($\Gamma = 1.85\pm0.26$, 15-150 keV
fluence = $(2.4\pm0.3)\times10^{-6}$ erg cm$^{-2}$, \citealt{GCN5732})
to the $1-2$ keV energy range, we obtain a fluence of
$(2.5\pm0.3)\times10^{-7}$ erg cm$^{-2}$, in agreement with our
estimate. \\
\qquad We thus conclude that, for this GRB, the prompt emission
at soft X--ray energies ($0.3-4$ keV) is consistent with an
extrapolation of the hard X--ray emission ($15-150$ keV),
suggesting that it probably has the same origin.
If we instead adopt the optical depth-energy relation in Eq.\ref{eq:predehl} we obtain a flux in the 1-2 keV band that is $\sim 3$ times greater than the BAT-extrapolated value. However, owing to the errors in the estimate of the spectral parameters of the ring and in the optical extinction $A_V$, this result is not conclusive.

\section{GRB 070129}

\begin{figure}[tb]
    \centering
    \includegraphics[width=0.25\textwidth]{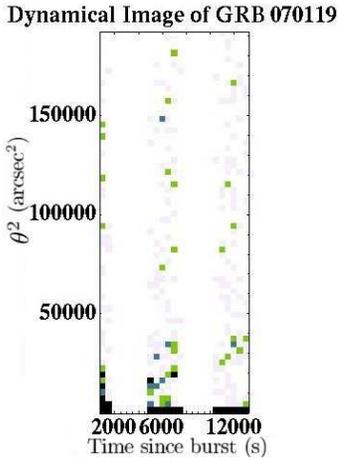}
    \caption{Dynamical image of \object{GRB 070129}. The  halo is visible
    in the second and third orbits. Is not visible in the first one due to the high background level and short
exposure time.}
    \label{fig:ddp}
\end{figure}

On 2007-01-29 the BAT instrument was triggered at 23:35:10 UT by a
precursor of \object{GRB 070129}, and the Swift satellite automatically
slewed to the derived sky position. This allowed detection of the
main peak of emission, which was detected at 23:41:30 UT by both the
BAT and XRT instruments. Despite the burst location at high
Galactic latitude (l=$157.2^\circ$, b=$-44.7^\circ$), in a region
of moderate absorption (optical extinction $A_{V}\simeq0.4$,
\citealt{Schlegel}), a detailed analysis of the XRT data led to the
discovery of a faint partial ring \citep{GCN6065} centered at the
afterglow position. The ring radius increased from $\sim1'$ to
$\sim3'$  during the second and third Swift orbits (from $\sim
T_{0}+4450$ s to $\sim T_{0}+13100$ s).
Figure \ref{fig:ddp} shows the dynamical image
obtained from the XRT Photon Counting data. The expanding ring is
evident in the second and third orbits. The cross section of the scattering process decreases rapidly with energy (see Eqs.\ref{eq:draine} and \ref{eq:predehl}) and becomes vanishingly small above a few keV,
so just as for the previous burst, we restricted our analysis to
energies below 4 keV.
From an annulus with radii of $5'$ and $8'$ centered on the
afterglow and excluding point sources and flickering pixels, we
estimated a background count rate of  $(8.9\pm0.7)\times 10^{-8}$
cts s$^{-1}$ arcsec$^{-2}$. Inspecting the dynamical image, we
selected the region containing the halo as an annulus of inner
radius of $90''$  and outer radius of $160''$. If we consider for the moment only the second orbit, this region
contains $45\pm6$ counts, of which $11\pm3$ are expected to be
background. Thus the net number of halo counts in the second orbit is $32\pm8$,
corresponding to a detection at the $\sim4\sigma$ confidence
level.

\begin{figure}[tb]
    \centering
    \includegraphics[width=0.4\textwidth]{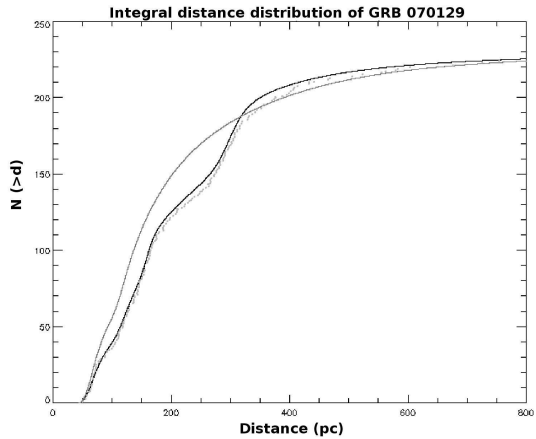}
    \caption{Integral dust distance distribution of \object{GRB 070129} (light gray points).
    The black line is  the simulated distribution with the two scattering rings
    at $D=150$ and $D=290$. The dark gray line is the distribution expected from
    a uniform background.
    }
    \label{fig:int_distr}
\end{figure}

Owing to the small number of halo counts, we based our analysis on
the integral distribution of $D$. This approach has the advantage
of using all the available information  and avoids the
arbitrariness of data binning.
The integral distributions of the Lorentzian peaks are arctan
profiles, with $D_{dust}$ corresponding to an inflection point.
Consider the expected halo region, which is an annulus centered on
the GRB position with inner radius of $\sim30''$  (in order to
exclude the afterglow) and outer radius of $\sim300''$  (the scattering cross section is negligible at larger
radii): the integral
dust distance distribution obtained from such a region using only
the second and the third orbits is reported in
Fig.~\ref{fig:int_distr}.
In order to model this distribution, we used the XRT observation of
the prompt X--ray emission as input for our simulation, 
initially with the optical depth-energy relation in Eq.\ref{eq:draine} and the background rate estimated
above. Then we  defined a grid of possible values for the distance
of the dust layer, and for every value of the grid we made a large
number of simulation runs in order to derive the expected shape
for the integral distribution. We then used the
Kolmogorov-Smirnov (KS) test to compare the observed distribution with
those generated by the simulation. We found that the model that described the observations (with a KS-test probability of
$55\%$) best corresponds to a dust distance of $D=290$ pc. In contrast, if we try to describe the observed distribution with
what is expected only from the background, we obtain a probability of
only $5\times10^{-5}$ (see Fig.~\ref{fig:int_distr}). We applied
the same procedure to a late observation of the afterglow of
\object{GRB 070129}, when the dust ring has disappeared. In this case the
distance distribution measured from the same annular region is described well (KS-test probability of $99\%$) by the one expected
from uniform background, demonstrating the accuracy of the
background distributions derived with our simulation. Inspecting
Fig.~\ref{fig:int_distr}, one can see that there is another
inflection point at $\sim$150 pc. This could be due to another, fainter dust ring. Using the same procedure as described above, we
found a significant improvement (KS-test probability of $90\%$)
with two layers of dust, at distances $D=150$ pc and $D=290$ pc.
We thus conclude that the halo observed around \object{GRB 070129} is
probably due to two dust layers along its line of sight.
The knowledge of the prompt X--ray emission should 
allow us to test and possibly distinguish between different proposed optical depth-energy relations for the scattering process like Eqs.\ref{eq:draine} and Eq.\ref{eq:predehl}, corresponding to different dust properties, such as grain size distribution and composition.
Our simulations with the best model derived above for the dust
distributions predicts $40\pm7$ halo counts in the second XRT
orbit using the Draine optical depth-energy relation (Eq. \ref{eq:draine}) and $18\pm4$ halo counts using
the Predehl relation (Eq.\ref{eq:predehl}). Although the first value is closer to the
observed one ($32\pm8$ cts), the inadequate statistics do not allow us to derive a firm
conclusion in favor of one of the two possibilities.

\begin{figure}[tb]
    \centering
    \includegraphics[width=0.4\textwidth]{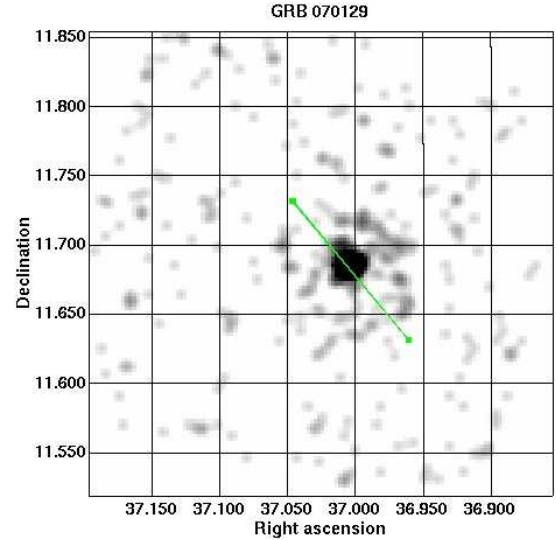}
    \caption{Image of the afterglow of \object{GRB 070129} obtained with XRT during the second
    Swift orbit since trigger time. The image has been rebinned and smoothed
    to show the faint diffuse structures better.
    The dust scattering ring is visible
    around the afterglow. The ring appears asymmetric, since is brighter in
    the region north west of the marked line. }
    \label{fig:2ndorbit}
\end{figure}

There is some evidence that the ring spatial distribution is not
uniform around the GRB position, being more intense in the
north west part. Dividing the halo region into two parts, as
indicated by the line in  Fig.~\ref{fig:2ndorbit}, we find
$24\pm5$ counts (halo+background) in  the north west part and
$10\pm3$ on the opposite side. We checked that this uneven
distribution, corresponding to a chance probability of
$\sim$1.6\%, is not due to a non uniformity in the XRT exposure.
The asymmetry in the ring intensity is thus probably due to a inhomogeneous distribution of the dust in the distant layer.

\section{Discussion}

The first dust scattering ring around  a GRB \citep{vaughan04} was
discovered thanks to the rapid localization provided by INTEGRAL
\citep{ibas} and the large collecting area at the X--ray energy of the
EPIC instrument on XMM-Newton. The results reported here
demonstrate that, despite its smaller collecting area, Swift can
also detect GRB halos, thanks to its capability to point in the
GRB direction within a few tens of seconds, thus catching the
possible rings near the maximum of their intensity \citep{proc06}.

Besides providing in
some cases useful information on the GRB prompt X--ray emission, detection of these expanding dust rings 
offers a powerful tool for mapping the three-dimensional distribution
of Galactic dust clouds. It should be noted that the derived
distances, based purely on dynamical arguments, can be extremely
accurate.

\begin{figure}[tb]
    \centering
    \includegraphics[width=0.4\textwidth]{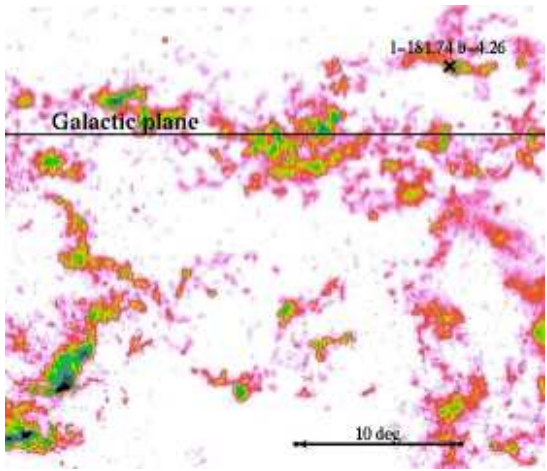}
    \caption{This $^{12}$CO (J=1-0) map \citep{dame} shows the molecular clouds present around the coordinates of the burst (marked by the cross). Both the burst position and the distance to the dust-scattering layer suggest identification of the layer with the molecular cloud \object{[KOY98] 66} \citep{kawamura}. The black line marks the position of the galactic plane.}
    \label{fig:mol_clouds}
\end{figure}
\begin{figure}[tb]
    \centering
    \includegraphics[width=0.4\textwidth]{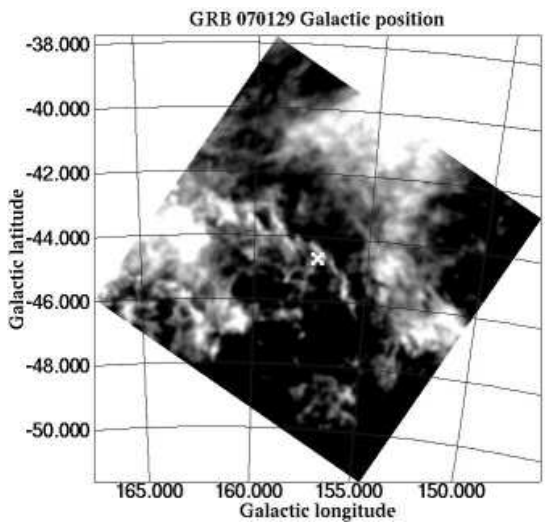}
    \caption{$A_V$ map \citep{Schlegel} around the \object{GRB 070129} position
(indicated by the cross).}
    \label{fig:av_map}
\end{figure}
This is exemplified by our results for \object{GRB 061019}. The line of sight of the burst passes through the Perseus spiral arm and near the galactic anti-center, where the usual kinematic methods of estimating distances are not applicable. The molecular clouds and nebulae present in this region (see Fig.\ref{fig:mol_clouds}) have been extensively studied \citep{lynds,dame,lee,kawamura}. In particular, the position of \object{GRB 061019} is consistent with the coordinates of the cloud \object{[KOY98] 66} of the $^{13}$CO (J=1-0) survey by  \cite{kawamura}, placed at a distance of $\sim 1.1$ kpc from Earth. Thus our determination of the distance to the dust layer ($D=941^{+43}_{-47}$ pc) suggests an identification with the molecular cloud \object{[KOY98] 66}, confirming and refining its distance estimate.\\
Assuming dust properties, one can extract some information about the
prompt X--ray emission of the GRB from the
spectrum and intensity of the halo, which is generally difficult to
observe directly. This is exemplified by our study of the \object{GRB
061019} halo: adopting the optical depth-energy relation in Eq. \ref{eq:draine}, we found a prompt
X--ray spectrum consistent with the extrapolation to the few keV
range of the prompt hard X--ray emission observed by Swift-BAT
\citep{GCN5732}. Note that this apparently unsurprising result is
not the rule, as shown by the case of GRB 031203
\citep{vaughan04,tm}, for which, independent of the adopted
dust models, the derived soft X--ray emission is not
compatible with an extrapolation of the hard X--ray spectrum.\\
When the prompt X--ray emission is known, the study of the rings
in principle permits us to verify and eventually distinguish between different
proposed scattering optical depth-energy relations (like Eqs.\ref{eq:draine} and \ref{eq:predehl}), which depends via scattering cross section on the dust properties
like size and composition. This is exemplified in our study of the
rings around \object{GRB 070129}, which could not however distinguish
between the two proposed formula, due to the inadequate
statistics of this halo.  Future observations of prompt X--ray
emissions followed by dust scattering rings as intense as those of
\object{GRB 061019} or GRB 031203 might lead to more concrete possibilities
of obtaining useful information on the properties of the interstellar
dust in  our Galaxy.

\begin{acknowledgements}

This research has been supported by the Italian Space
Agency through contract ASI-INAF I/023/05/0.

\end{acknowledgements}

\bibliographystyle{aa}
\end{document}